# Tuning the magneto-electrical properties of multiferroic multilayers through interface strain and disorder.

J. Gonzalez Sutter[1,2], A. Sarmiento Chávez[1,2], S. Soria[3], M. Granada[1,2], L. Neñer[1,2], S. Bengió[4], P. Granell[5], F. Golmar[6], N. Haberkorn[1,2], A. G. Leyva[7], M. Sirena[1,2].

**AFFILIATIONS**

[1]Instituto de Nanociencia y Nanotecnología, CNEA – CONICET, San Carlos de Bariloche, Río Negro, 8400, ARGENTINA.
[2]Instituto Balseiro, CNEA &UNCuyo, San Carlos de Bariloche, Río Negro, 8400, Argentina.
[3]División deCiencia de Materiales, Centro Atómico Bariloche, San Carlos de Bariloche 8400, Argentina.
[4]División deFísica de Superficies, Centro Atómico Bariloche, San Carlos de Bariloche 8400, Argentina.
[5]Instituto Nacional de Tecnologia Industrial, San Martín, Buenos Aires, Argentina
[6]ECyT – CONICET, Universidad de San Martín, Av. 25 de Mayo y Francia, San Martín 1650,Argentina.
[7]Departamento de Materia Condensada, Centro Atómico Constituyentes,Buenos AiresB1650, Argentina.

a)Authors to whom correspondence should be addressed: sirena@cab.cnea.gov.ar

**ABSTRACT**

Artificially engineered superlattices were designed and fabricated to induce different growth mechanisms and structural characteristics. DC sputtering was used to grow ferromagnetic ($La_{0.8}Ba_{0.2}MnO_3$) / ferroelectric ($Ba_{0.25}Sr_{0.75}TiO_3 or BaTiO_3$) superlattices. We systematically modified the thickness of the ferromagnetic layer to analyze dimensional and structural effects on the superlattices with different structural characteristics. The crystalline structure was characterized by X-Ray diffraction and transmission electron microscopy. The magnetic and electronic properties were investigated by SQUID magnetometry and resistance measurements. The results show that both strain and structural disorder can significantly affect the physical properties of the systems. Compressive strain tends to increase the competition between the magnetic interactions decreasing the ferromagnetism of the samples and the localization of the charge carrier through the electron-phonon interaction. Tensile strain reduces the charge carrier localization, increasing the ferromagnetic transition temperature. Structural defects have a stronger influence on the magnetic properties than on the transport properties, reducing the ferromagnetic transition temperature while increasing the magnetic hardness of the superlattices. These results help to further understand the role of strain and interface effects in the magnetic and transport properties of manganite based multiferroic systems.

## I. Introduction.

In 1988,Albert Fert and Peter Gruenberg discovered the giant magnetoresistance effect [1,2], giving birth tospintronics, a new research area in Physics. This started the search for new system and devices that could combine the electronic and the magnetic properties of charge carriers renewing the interest in manganites systems. Their colossal magnetoresistance effect, high Curie temperature (Tc) and half-metallic nature made them excellent candidates for the fabrication of spintronics devices. Nevertheless, earlier works showed in manganite tunnel junctions, a dramatic loss of the magnetoresistive effect well below the Tc of the electrodes, highlighting the influence of interface effects in these compounds [3]. Indeed, the interplay between the structural, magnetic and transport properties in manganites was the origin of this and many other phenomena. This boosted research of strongly correlated materials for surface engineering and system functionalization. New properties and functionalities emerge from combining different oxide compounds in multilayered structures, e.g. 2DEG [4,5], exchange bias [6,7,8], and superconductivity [9], among others. In this context, multiferroic multilayered systems combining ferromagnetic and ferroelectric oxides have being studied aiming to control the magnetization of spintronics devices using electrics fields [10,11]and searching for increased functionality[12,13].Manganite based multiferroic systems gave the opportunity to study many interesting phenomena in fundamental physics, i.e. magneto-electric coupling [14,15], ferroelectric effects [16-18] and in applied physics, i.e. resistive switching [19,20] and ferroelectric tunneling [21].

As mentioned before, strain and interface effects in these systems are critical for the development of new applications and relevant work has been done regarding this issue [22-24]. Nevertheless, further efforts are needed to understand the fundamentals behind the complex relationship between structural, magnetic and transport properties in these systems. Understanding and controlling conducting–ferromagnetic/insulator–ferroelectric interfaces is crucial to develop multiferroic heterostructures for new technological applications. We have studied the size effects on the magnetic and the transport properties of artificially engineered manganite-based multiferroic superlattices. We have fabricated multiferroic superlattices (SL) composed of ferromagnetic (FM) $La_{0.8}Ba_{0.2}MnO_3$ (LBMO), ferroelectric (FE) $Ba_{0.25}Sr_{0.75}TiO_3$ (BSTO) and $BaTiO_3$ (BTO) compounds. Playing with the lattice mismatch of the composing layers it is possible to tune the structural growth of the hetero-structures, choosing to enhace the effects of strain or structural defects effects. LBMO presents nominally the same lattice parameters BSTO (0.392 nm), while LBMO and BTO (~0.4 nm) present an important lattice mismatch.

## II. Experimental Details.

Coherently strained LBMO/BSTO and disordered LBMO/BTO superlattices (SL) were fabricated by DC sputtering. Samples were grown from stoichiometric ceramic targets over single crystalline SrTiO3 (100) substrates and oxygenated to reduce the influence of oxygen vacancies on the physical properties of the systems. Details of the deposition procedure are given in reference [25]. Two series of samples were grown varying the thickness of the ferromagnetic layer (2nm<$t_{FM}$<16nm) while keeping the thickness of the ferroelectric layer constant (5nm). This was done in order to study the dimensional effect, under different stress and disorder characteristics, on the physical properties of the nanostructures. As the thickness of the ferromagnetic layer is reduced, an increasing influence of the interface effects is expected. The number of bilayers in the samples was chosen to keep the total thickness of the SL as close to 120 nm as possible. This was done in order to reduce size effects in the morphological properties of the samples.

The structural and surface properties of the superlattices were studied by standard θ-2θ X-ray diffraction (XRD), atomic force microscopy (AFM) and transmission electron microscopy (TEM). Rocking curves were performed around the central diffraction peak of the superlattices and full width at half maximum (FWHM) values were analyzed to obtain information about the dispersion of the crystalline structure around the main diffraction orientation, i.e. structural disorder. A commercial superconducting quantum interference device magnetometer (SQUID) was used to study the magnetic properties of the samples for magnetic fields (H) up to 5 T and for temperatures (T) between 4 K and 400 K. Standard four probe configuration was used to study the transport properties of the superlattices using magnetic fields up to 1 Tesla for temperatures between 4.25 K and 300K.

### III. Results.

### III.1. Structural properties.

Figure 1 presents the X-ray patterns of the LBMO/BSTO and LBMO/BTO superlattices with different thicknesses of the ferromagnetic layers. The patterns show a textured growth of the samples in the *c* direction perpendicular to the sample surface, with [00*l*] diffraction peaks and the presence of satellite peaks due to the superstructure. This is a good indicator of the quality of the superlattices structure. The thickness of the composing FM/FE bilayers was obtained from the peak to peak distances in XRD patterns for the different samples. Due to the similarity of the LBMO and the BSTO lattice parameters, the center peak of the SL diffraction pattern is a good indication of the lattice parameter of the SL. From this value the in-plane lattice parameters and their difference respect the bulk compound can be estimated (i.e. SL stress). For the BSTO based SL, a systematic shift of the central peak towards lower diffraction angles is observed for decreasing ferromagnetic layer thicknesses. Typically compressive and tensile stresses smaller than 0.3% of the bulk lattice parameter were obtained. FWHM of the rocking curves for the BTO SL can go from 0.2° to 1.2°, while the FWHM of a crystalline $SrTiO_3$ substrate is around 0.07°.

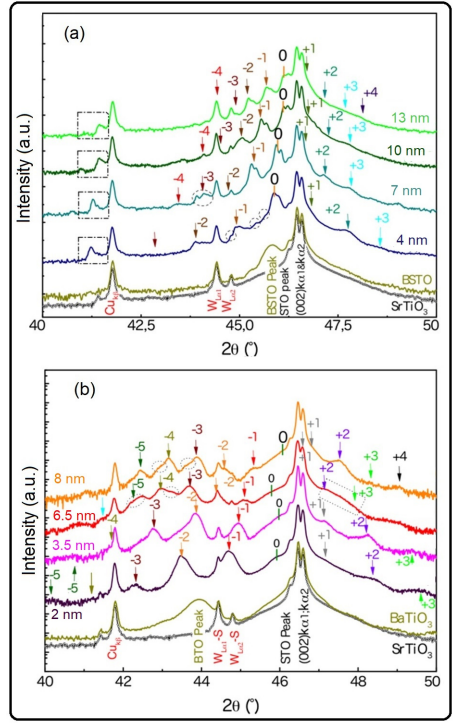

Figure 1: X-ray profiles for the BSTO (a) and BTO (b) superlattices with different thicknesses of the ferromagnetic layers. The arrows indicate the position of the satellite peaks. The contribution of the $SrTiO_3$ substrate is also indicated.

Being *a* and *b* the directions in the plane of the SL, and assuming the stress $\varepsilon_{aa} = \varepsilon_{bb}$, the stress component in the c direction can be calculated as [26]:

$$\varepsilon_{cc} = - 2\nu / (1-\nu)\, \varepsilon_{aa}. \qquad (1)$$

Where $\nu$ is the Poisson's ratio. In a first-order approximation $\nu$ can be considered to be 0.5, corresponding to a unit cell structure deformation, with volume conservation. From the *c* axis lattice parameter obtained from the XRD patters, and the *c* axis lattice parameter of the LBMO target (bulk sample), the in-plane stress can be calculated. Figure 2 displays the stress in the BSTO based SL for different thicknesses of the FM layer.

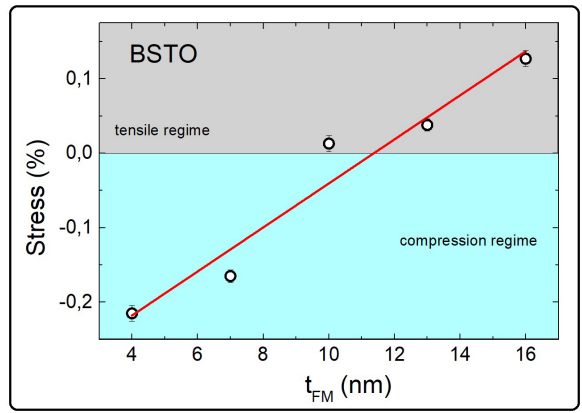

Figure 2: Stress in coherently grown superlattices (BSTO based SL) as a function of the thickness of the ferromagnetic layer.

BSTO based SL with small values of $t_{FM}$ ($t_{FM}$<10 nm) present a small compressive stress. SrTiO3 substrates have a smaller lattice parameter (0.3905 nm) than LBMO and BSTO, inducing the small compressive strain observed for these samples. As the thickness of the FM layer increases, strain in the sample relaxes and even a small tensile strain is calculated for samples with a thicker LBMO layer ($t_{FM}$> 12 nm).

Figure 3 shows a typical AFM image of the manganite based SL. Surface roughness is around 2 u.c with a particle density, i.e. outgrowth defects, of approximately 0.1 def/$\mu m^2$. These values are in agreement with the values obtained in similar systems [18]. Since surface defects in manganite electrodes tend to shortcut tunnel junctions like devices, a low density of surface defects is desired in these systems. In this context, the cleanness of the substrate is more relevant than the SL composition. On the other hand, no systematic behavior of the SL roughness with the FM layer was found.

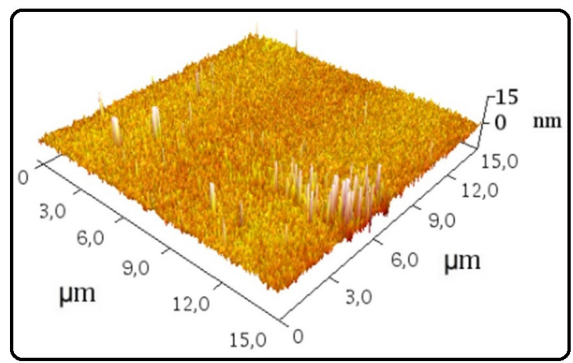

Figure 3: Typical AFM image of manganite base superlattices, shown for a BSTO SL with a thickness of the FM layer of 10 nm.

Figure 4 presents the TEM images of BSTO (a) and BTO (b) based superlattices. The images show a multilayer structure in agreement with the XRD patterns. TEM images of BSTO based superlattices show clean and sharp interfaces, atoms in the structure and the crystal growth direction are well defined with little deviations. Images are consistent with a coherent growth of the samples due to the small lattice mismatch between the SrTiO$_3$ substrate, the LBMO layer and the BSTO layer. On the other hand, BTO based SL shows an important presence of dark areas spatially correlated. These areas could be originated in structural defects and strain fields in the sample. Indeed, this is in agreement with the higher values of FWHM obtained from the rocking curves of these samples compared with the BSTO based SL. Higher FWHM values for the rocking curve indicate a higher dispersion of the crystalline directions around the main crystal orientation. These results indicate high density of structural defects in the BTO based superlattices, compared with BSTO multilayers.

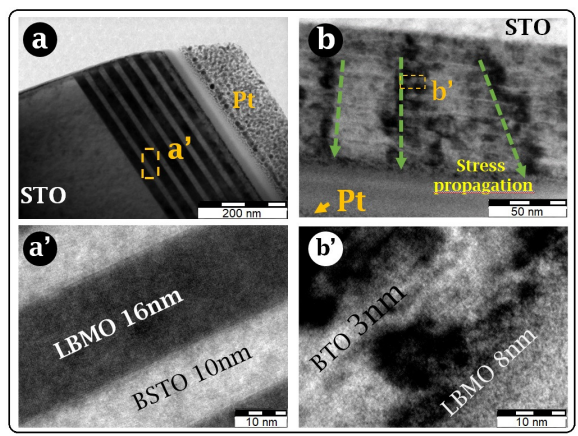

Figure 4: TEM images of BSTO based superlattices (a) and BTO based superlattices (b). The bottom panels are a close in of the chosen area in each image.

### III.2 Magnetic properties.

Figure 5 presents the temperature dependence of the magnetization (M) of BSTO based SL (a) and BTO based SL (b). Magnetization in the samples was normalized by the total FM volume in each sample. LBMO layers in SL present high Curie Temperatures (Tc) going from a high temperature paramagnetic state to a low temperature ferromagnetic regime, confirmed by magnetization vs. magnetic field measurements (not shown). The magnetic ordering temperature strongly depends on the thickness of the FM layer in the structure.

The Tc decreases as the FM layer thickness decreases. This effect is generally ascribed to an increasing competition between the double exchange (ferromagnetic) interaction and the superexchange (antiferromagnetic) interaction. Both compressive strain and structural disorder in the sample decrease the double exchange interaction through changes in the Mn-O bond and increased concentration of ion vacancies that localize the electrons in the structure [23,24,27]. This effect is the origin of the magnetization reduction also observed in the samples as the thickness of the FM layer is reduced. Superlattices with a thickness of the FM layer smaller than a critical value show no FM transition as a function of temperature. This critical thickness seems to be around 2.5 nm for the BSTO based SL and 2 nm for the BTO based ones. This is consistent with the values in the literature [27,28,29] and it is related to the presence of structural defects, chemical and charge interdiffusion at the interfaces. The BTO based multilayer with a FM thickness of 3.5 nm seems to present an anomalous behavior probably due to problems during the field cooling of the sample.

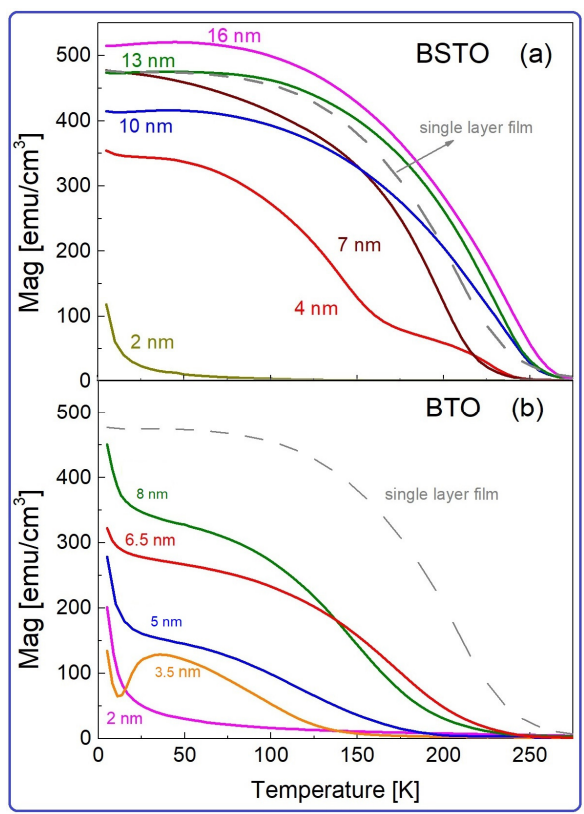

Figure 5: Temperature dependence of the magnetization of BSTO based SL (a) and BTO based SL (b). Magnetization was measured with an applied field of 300 Oe and was normalized by the total FM volume

in each sample. The magnetization vs. temperature of a single manganite layer film with a thickness of 96 nm is included as a reference.

For BSTO SL with thicker FM layers ($t_{FM}$ > 10 nm), the magnetic transition temperature is higher than the one observed for the reference single layer film. This is consistent with the stress regime calculated from the XRD patterns previously shown in Figure 2. SL with thin FM layers present a compressive strain that reduce the double exchange mechanism. However, when the thickness of the ferromagnetic layer is higher than 10 nm a small tensile strain appears in the sample. This tensile strain straightens the Mn-O bonds, increasing the orbital degeneracy and reducing the Jahn-Teller distortion present in these systems [30]. All these effects increase the charge delocalization and the double exchange mechanism, increasing the ferromagnetism of the sample. For equivalent thicknesses of the FM layer, BTO based SL present lower values of magnetization and lower magnetic transition temperatures. This seems to indicate that the structural disorder at the interface, induced by the lattice mismatch between the BTO and LBMO has a greater impact on the magnetic properties of the SL than the strain induced in the BSTO multilayers.

Figure 6 shows Tc, defined as the maximum of 1/M * dM/dT, as function of the thickness of the FM layer in the BSTO and BTO based SL. For increasing thickness of the FM layer in the SL, Tc increases and seems to become higher than the magnetic transition temperature observed for the single layer reference film with a thickness of 96 nm. The Tc is reduced for a thickness of the LBMO layer lower than 6 nm. As mentioned before the Tc in BTO based SL seems to be lower than the one observed for BSTO SL with equivalent thickness of the FM layers

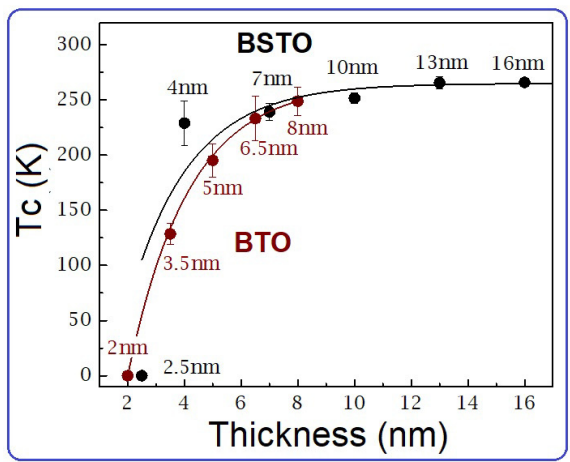

Figure 6: Magnetic transition temperature as a function of the thickness of the ferromagnetic layer in the superlattices structure.

.

For thicknesses of the FM layer lower than 4 nm there is a rapid drop of the magnetic transition temperature for decreasing thickness of the FM layers. This is related with the increasing weight of interface effects in these samples. Chemical, structural and magnetic disorder in the interface increase the competition between the different magnetic interactions [31], favoring the formation of antiferromagnetic zones [6,7]

Millis et. al. studied in 1998 the relationship between the magnetic transition temperature and strains (ε) in manganite films, considering the electron-phonon interaction and the double exchange hopping [32]. It was found that:

$$Tc(\varepsilon) = Tc(0) + 0.5\Delta_T\, \varepsilon^2 \quad \text{where} \quad \Delta_T = \frac{\partial^2 Tc}{\partial \varepsilon^2} \qquad (2)$$

As indicated by Angeloni et. al., this leads to an increase of the transition temperature for tensile strains and a decrease of Tc with compressive strains [30]. Figure 7 shows the magnetic transition temperature as a function of in plane stress for the BSTO based superlattices.

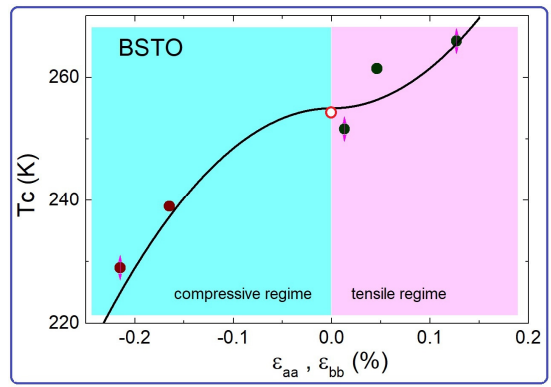

Figure 7: Magnetic transition temperature as a function of the biaxial strain in BSTO based superlattices. The line is a fit using equation 2 (see text). The Tc of a "thick" single layer film is included as a reference.

The dependence of Tc with in-plane stress seems to be consistent with equation 2. A reduction of Tc with increasing compressive strain is observed, while an increase of Tc is observed for increasing tensile strain. The value of Tc, obtained for unstrained samples is the same than the value measured for the reference single layer film, and to the saturation value observed for BTO based SL. For these systems, the biaxial strain is expected to relax through structural disorder at the interface. From the fitting of the data a value of 1300 K is obtained for $\Delta_T$. This large value is consistent with the value obtained for a similar system [33], and was ascribed to the strong interaction between the charge carries and the localizing effect of a strong lattice coupling, i.e. Jahn-Teller effect. The large value of $\Delta_T$ is an indication of the strong effects of strain on the magnetic and transport properties of manganite based systems.

BSTO and BTO based SL with very thin ferromagnetic layers ($t_{FM}$<3nm) presented a paramagnetic behavior in all the measured temperature range. Samples with thicker ferromagnetic layers, presented the typical hysteresis loops for the *M(H)* curves, for temperatures lower than the Tc of the samples. Figure 8 presents the coercive field $H_c$ as a function of the FM thickness in the BSTO and BTO based SL, measured at 5 K.

For both series of samples, with high values of the FM thickness ($t_{FM}>5$ nm), the coercive field decreases with increasing LBMO thicknesses. Indeed, the samples seem to present a linear dependence of the coercive field with the inverse of the FM thickness. This is the expected behavior for a simple surface contribution and it was observed before for single manganite films [34]. Nevertheless, it should be noted that Monsen et. al. reported a $1/t_{FM}^2$ dependence of the coercive field of single manganite films grown over SrTiO$_3$ substrates, suggesting a more complex behavior than the one expected from a simple surface contribution.

BTO based SL present higher coercive fields compared with BSTO SL. This indicates that structural disorder is more efficient as a pinning mechanism in manganite films that coherent strain. Structural defects probably increase the density of pinning centers at the interface. Equivalent results were obtained for single La$_{0.6}$Sr$_{0.4}$MnO$_3$ single films grown over MgO and SrTiO$_3$ substrates [34]

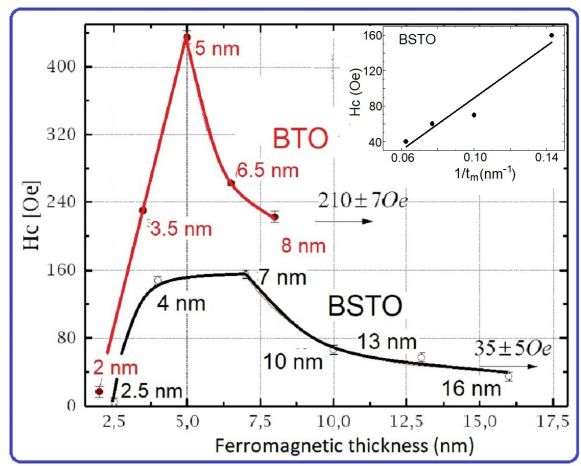

Figure 8: Coercive field measured at 5 K as a function of the thickness of the ferromagnetic layer in the BSTO and BTO based superlattices. Lines are guides for the eyes. The inset shows the linear dependence of the coercive field as function of the inverse of the ferromagnetic layer thickness.

For low thicknesses of the FM layer ($t_{FM}<5$nm) the coercive field decreases for decreasing thickness. This is probably related, as mentioned before, with the decreasing ferromagnetism of the samples due to an increasing competition between the different magnetic interactions.

III. 3 Transport properties.

Figure 9 shows the temperature dependence of the zero-field resistivity for the BSTO and BTO based superlattices. The temperature dependence of the resistivity for the SL is consistent with the magnetic measurements. The samples present a high metal-insulator transition temperature (Tp), with a high temperature paramagnetic/insulator state and a low temperature ferromagnetic/conducting state. Superlattices with a thickness of the ferromagnetic layer higher than a critical thickness show a high temperature metal-insulating transition with low resistivity values, close to the ones found for the single LBMO reference layer. This critical thickness is around 7 nm for BSTO based SL and 5 nm for BTO based multilayers. For decreasing thickness of the ferromagnetic layer, the resistivity of the sample increases and the metal-insulator transitions shifts to lower temperatures.

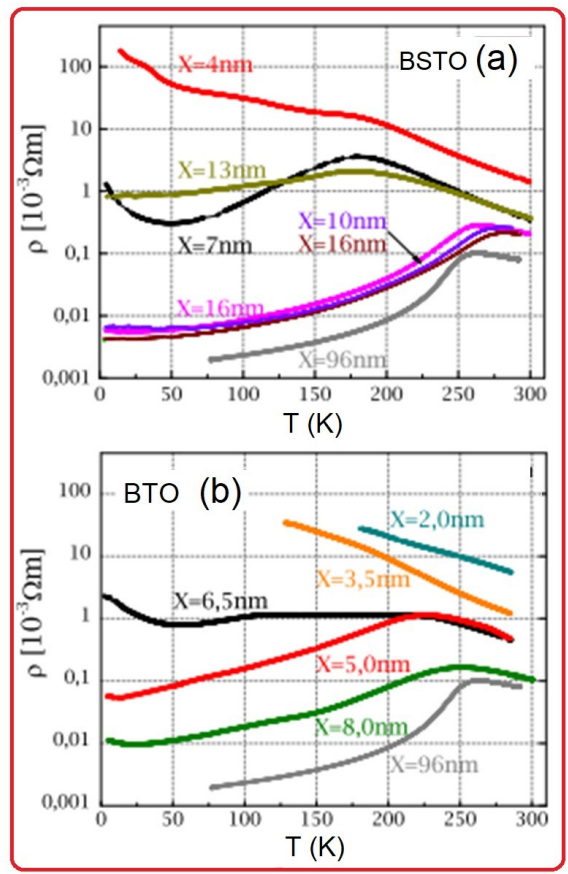

Figure 9: Temperature dependence of the resistivity for the BSTO (a) and BTO (b) superlattices, without an applied magnetic field.

Manganite based SL with a thickness of the ferromagnetic layer lower than the critical thickness show an increased resistivity and a reentrant semiconductor like behavior at low temperatures (T<<Tp). SL with very thin ferromagnetic layer ($t_{FM}$<4nm) show a semiconductor-like behavior in the entire temperature range. Compressive strain in BSTO based SL and interface structural disorder in BTO based SL, strongly affect the transport properties of these systems. According to the magnetization and resistivity measurements, the strong electron-phonon interaction localizes the charge carriers even in the ferromagnetic regime.

BTO based SL show a broad transition width, consistent with the reduced homogeneity of the samples. The samples present higher resistivity values than the reference "thick" single layer sample. This is a consequence of the interface structural disorder. Vacancies and dislocations break the Mn-O-Mn bonds weakening the double exchange mechanism responsible for the ferromagnetism and metallic behavior in these samples. Due to the stress relaxation at the interfaces, the middle of the LBMO layers is relatively free of strain. However, at low temperatures the disorder tends to localize the charge carriers.

The band structure of manganites is characterized by a localization energy (known as the "mobility edge") due to the characteristics of the electrostatic potential within the sample. Electronic states with energies lower than this mobility edge are localized while the ones with higher energy are conducting states. The energy difference between the mobility edge and the Fermi energy of the system determines the "pseudo gap" ($\Delta$) and its magnitude compared with the thermal energy, determines the conducting mechanism in the sample. At high temperatures, the main transport mechanism is a thermally activated process with the resistivity given by [35]:

$$\rho = \rho_0 \exp(-\Delta/K_B T) \qquad (Ec. 1)$$

being $\rho_0$ the expected resistivity of the system at very high temperatures. At low temperatures the main acting mechanism is the variable range hopping:

$$\rho = \rho_1 \exp(-Q/K_B T)^{1/4} \qquad (Ec\ 2)$$

where $\rho_1$ is a constant that depends on the electron-phonon interactions.

The strain field in manganite films modifies the band structure of the samples [30]. Transport measurements in BSTO SL seem to show that for thin FM layers the samples present a relevant compressive strain field that increases $\Delta$. This increases the localization of the charge carriers, requiring higher temperatures to delocalize them. A similar behavior was observed when decreasing the thickness of single layer manganite films [35].

Figure 10 shows the metal-insulator transition (Tp) as a function of the ferromagnetic layer's thickness for BSTO and BTO based SL. Consistently with the magnetic measurements, the metal-insulator transition decreases for decreasing thickness of the ferromagnetic layer for the two series of samples. As mentioned before, decreasing the $t_{FM}$ has two effects: an increasing influence of interfacial effects, but also an increasing influence of structural defects in the BTO based SL and the presence of compressive biaxial strains in the BSTO SL. For the latter, thicker FM layers ($t_{FM} > 10$ nm) induce a tensile strain in the sample that increases the ferromagnetic transition temperature and the metal-insulator transition is high. Indeed, Tp for the BSTO SL with a FM layer's thickness of 16 nm seems to be higher than the one measured for the single layer reference film.

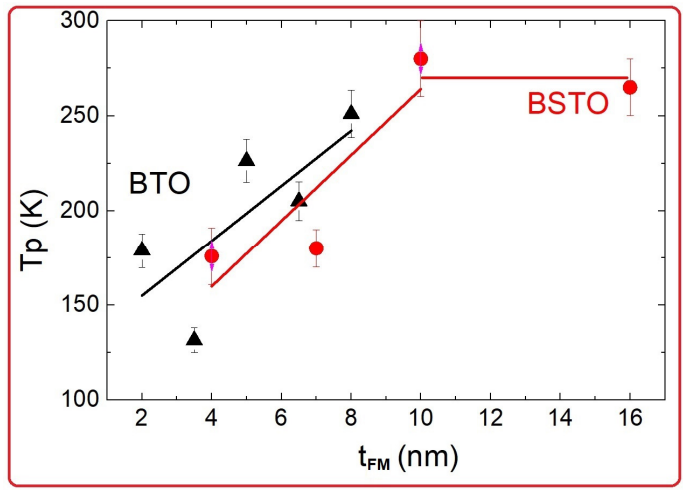

Figure 10: Metal insulator transition temperature as a function of the ferromagnetic thickness for BSTO and BTO based superlattices.

We have seen that the localization of the charge carriers plays a fundamental role in the transport properties of the SL. In the following, size effects on the localization of the charge carriers are analyzed. Figure 11 shows the typical temperature dependence of the resistivity at high temperatures (T>Tp) (a) and at low temperatures (T<<Tp) (b), shown for BSTO based SL.

The high temperature regime seems to be consistent with a thermally activated behavior, while the low temperature regime seems to be consistent with a variable range hopping behavior. This is in agreement with the results for single layer manganite films [35] and for similar systems in the literature [23, 24, 33].

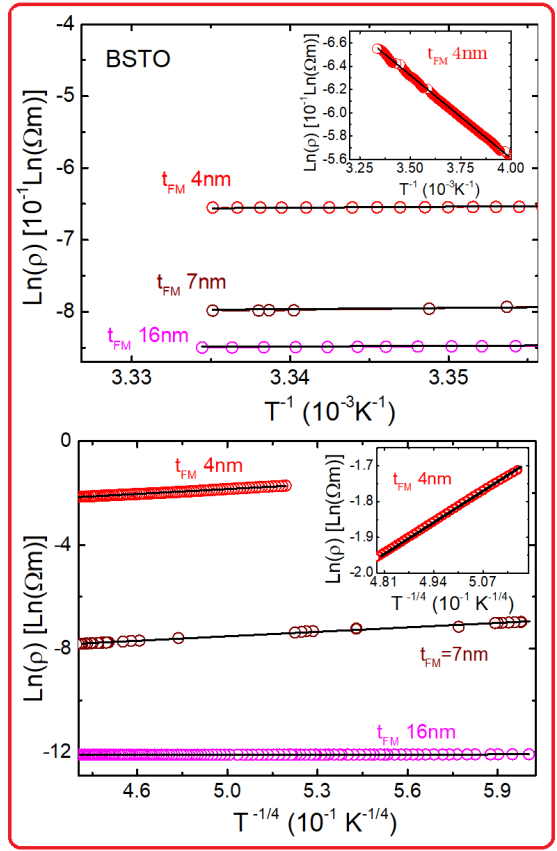

Figure 11: Typical temperature dependence of the resistivity in the high temperature regime (a) and the low temperature regime (b). Show for the BSTO based superlattices. The insets show a close in of the data fitting.

Figure 12 presents the thermal activation energy as a function of $T_{FM}$ for BSTO and BTO based superlattices. As the thickness of the ferromagnetic layer decreases, the difference between the mobility edge and the Fermi energy increases; indicating a greater localization of the charge carriers. This is consistent with an increase of structural defects that increase the localization energy and higher compressive strains that favor the localization of the charge carriers through electron-phonon interaction. These results are in agreement with the reduction of the ferromagnetism in these samples. BTO based SL present a lower activation energy than BSTO based SL, indicating a lower localization of the charge carriers. This is consistent with the idea that during the growth, strains relax by structural defects in the first nanometers and then the FM layer grows with a reduced influence of strain and defects. The activation energy of superlattices with "thick" ferromagnetic layers (~80 meV) is comparable with the values found for single layer manganite films in the literature [35].

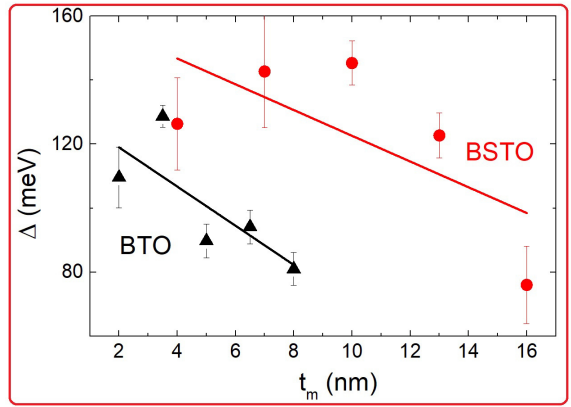

Figure 12: Activation energy of the thermally activated regime for the BSTO and BTO based superlattices. Lines are guides for the eyes. The data were obtained from linear fits to high-temperature resistivity as those presented in Fig. 11.

Figure 13 shows the low temperature activation energy, corresponding to the variable range hopping regime, as a function of $t_{FM}$, for both series of SL. Decreasing the thickness of the ferromagnetic layer in the superlattices, increases the activation energy in the variable range hopping regime. For BSTO based SL, below a critical thickness ($t_{FM}<10$ nm), the activation energy seems to increase linearly with decreasing $t_{FM}$.

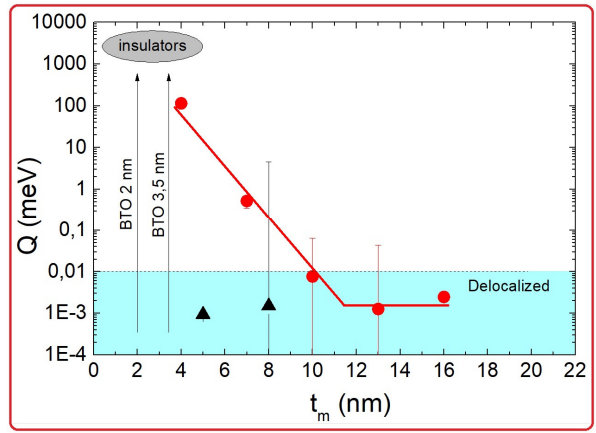

Figure 13: VRH energy as a function of the ferromagnetic thickness for BSTO and BTO superlattices. The data were obtained from linear fits to low-temperature resistivity as those presented in Fig. 11

For the BTO based SL and for thicknesses of the ferromagnetic layer higher than 4 nm, the charge carriers seem to be delocalized. Below this critical thickness, the samples present a semiconductor-like behavior in the entire temperature range. The low temperature regime was not available due to the high resistivity of these samples. These results seem to validate the idea that an important structural disorder is located at the layers interface. For thicknesses of the FM layer higher than this critical thickness, transport is mainly driven through the region relatively free of stress and disorder in the center of the layer. When the thickness is reduced, this is not possible and a huge change in the localization of the charge carriers is observed.

At low temperatures, the possibility for the charge carriers to "jump" between different localization volumes effectively reduces the energy required to delocalize the electrons. Indeed, the low temperature activation energy, $Q$, is reduced in more than one order of magnitude, compared with the high temperature activation energy, $\Delta$.

### IV. Conclusions.

Artificially engineered superlattices were grown to study the size effects on the structural, magnetic and transport properties of multiferroic multilayers under different growth conditions. We have found that both strain and structural disorder can affect the magnetic and transport properties of the system. Compressive strain tends to increase the competition between the magnetic interactions decreasing the ferromagnetism of the samples. Compressive strains increase the localization of the charge carrier through the electro-phonon interaction. On the other hand, tensile strain reduces the charge carrier localization, increasing the ferromagnetism in the sample. Structural defects at the interface introduced to relax the biaxial strain during the growth of the superlattices affect the magnetic and transport properties of the superlattices in a similar way. This kind of structural defects was found to strongly affect the magnetic properties while its influence on the transport properties is reduced. These results help to further understand the role of strain and interface effects in the magnetic and transport properties of manganite based multiferroic systems. A better understanding and control of interface effects is critical for the development of devices with new functionalities.

**AUTHOR'S CONTRIBUTION**
All authors' contributions were necessary for the realization of this work.

**ACKNOWLEDGEMENTS**

The authors acknowledge financial support from the "Agencia Nacional para la Promoción Científica y Tecnológica" **PICT 2018-01597** and **PICT 2018- 03126**. M. S. acknowledges financial support from EEC: PEOPLE MARIE CURIE ACTIONS, Research and Innovation Staff Exchange (RISE), Call: H2020-MSCA-RISE-2016 , Novel Magnetic Nanostructures for Medical Applications – **MAGNAMED**.

**AIP PUBLISHING DATA SHARING POLICY.**

The data that support the findings of this study are available within the article and additional data are available from the corresponding author upon reasonable request.

**REFERENCES.**

# FIGURE LEGENDS

Figure 1: X-ray profiles for the BSTO (a) and BTO (b) superlattices with different thicknesses of the ferromagnetic layers. The arrows indicate the position of the superlattices peaks. The contribution of the SrTiO$_3$ substrate is also indicated.

Figure 2: Stress in coherently grown superlattices (BSTO based SL) as a function of the thickness of the ferromagnetic layer.

Figure 3: Typical AFM image of manganite base superlattices, shown for a BSTO SL with a thickness of the FM layer of 10 nm.

Figure 4: TEM images of BSTO based superlaticess (a) and BTO based superlattices (b). The bottom panels are a close in of a chosen area in each image.

Figure 5: Temperature dependence of the magnetization of BSTO based SL (a) and BTO based SL (b). Magnetization was measured with an applied field of 300 Oe and was normalized by the total FM volume in each sample. As reference the magnetization vs. temperature of a single manganite layer film with a thickness of 96 nm is included.

Figure 6: Magnetic transition temperature as a function of the thickness of the ferromagnetic layer in the superlattices structure.

Figure 7: Magnetic transition temperature as a function of the biaxial strain in BSTO based superlattices. The line is a fit using equation 2 (see text). The Tc of a "thick" single layer film is included as a reference.

Figure 8: Temperature dependence of the resistivity for the BSTO (a) and BTO (b) superlattices, without an applied magnetic field.

Figure 9 Metal insulator transition temperature as a function of stress for the BSTO Superlattices (left) and as function of the X-Ray rocking curves FWHM for the BTO superlattices (right). Lines are guides for the eyes.

Figure 10: Typical temperature dependence of the resistivity in the high temperature regime (a) and the low temperature regime (b). The insets show a close in of the data fitting.

Figure 11: Activation energy of the thermally activated regime for the BSTO and BTO based superlattices. Lines are guides for the eyes.

Figure 12: VRH energy as a function of stress at the thermal activation regime (left panel) and at the low temperature variable range hopping regime (right panel) Lines are guides for the eyes.

Figure 1: Gonzalez Sutter, et. al.

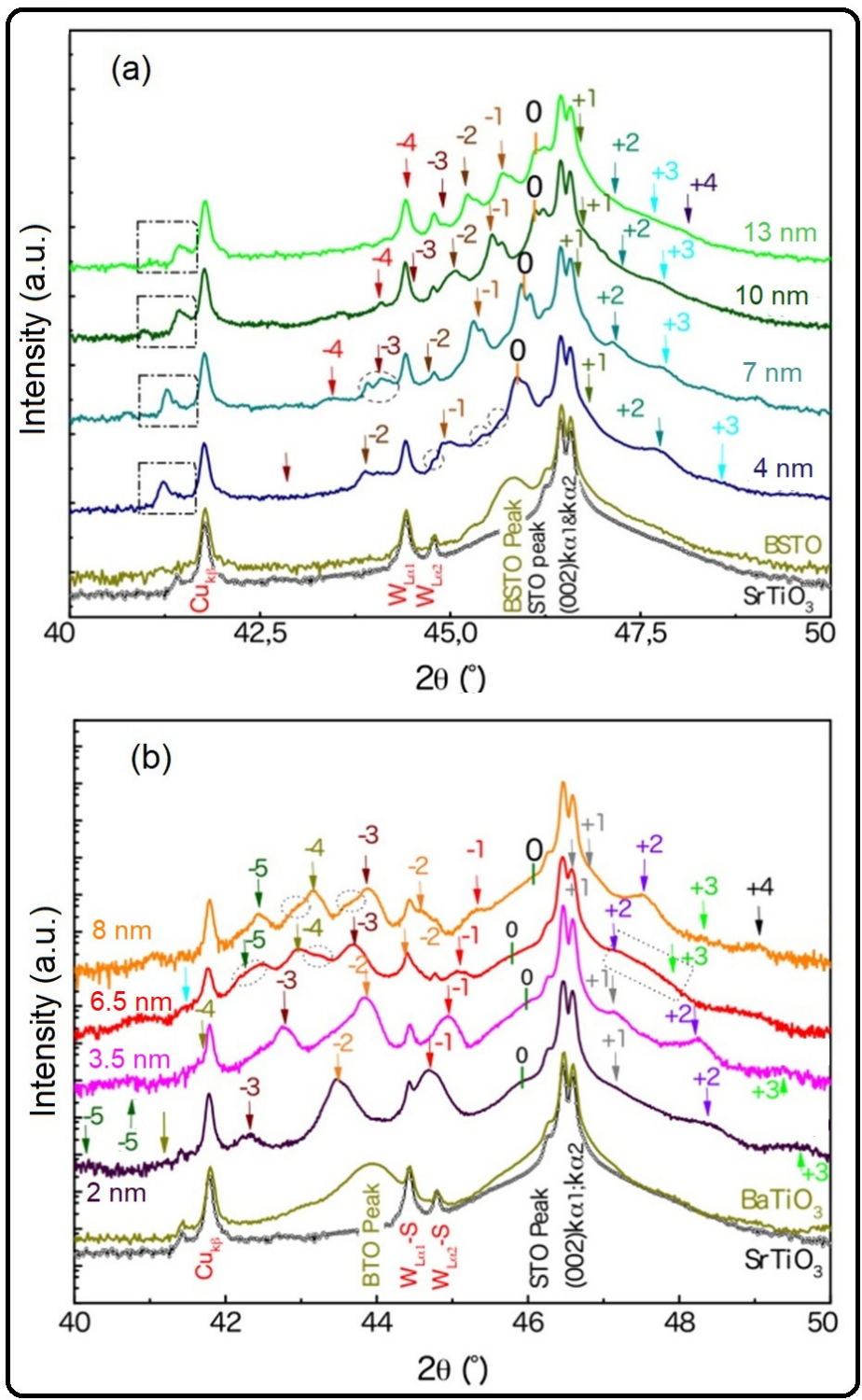

Figure 2: Gonzalez Sutter, et. al.

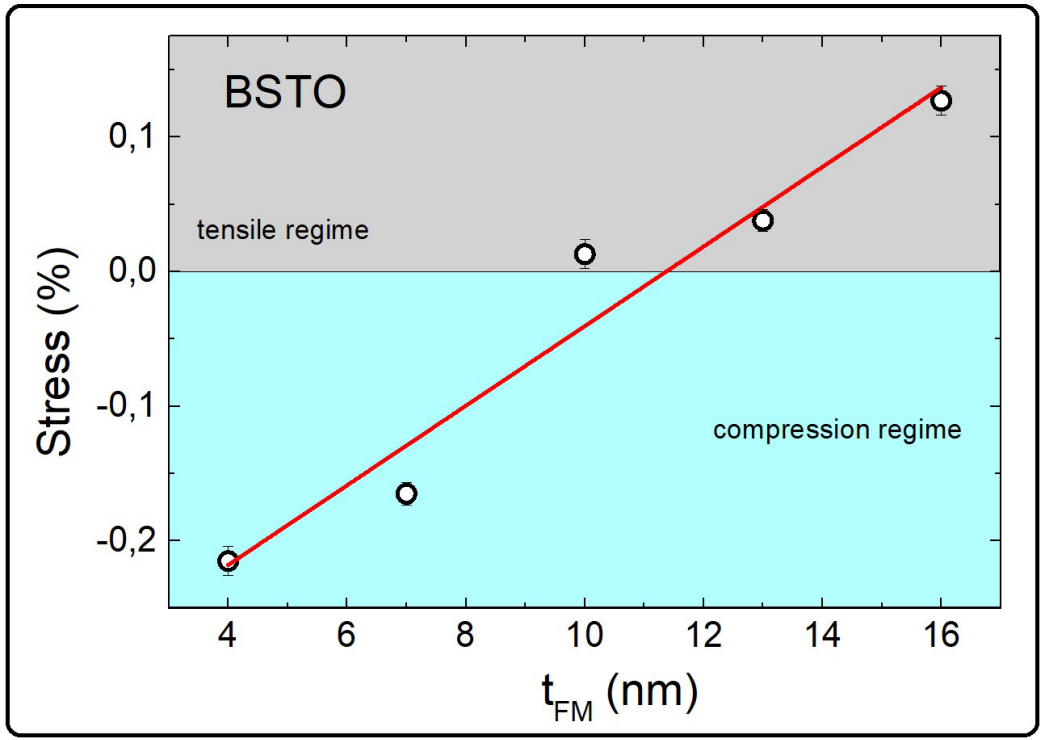

Figure 3:Gonzalez - Sutter, et. al.

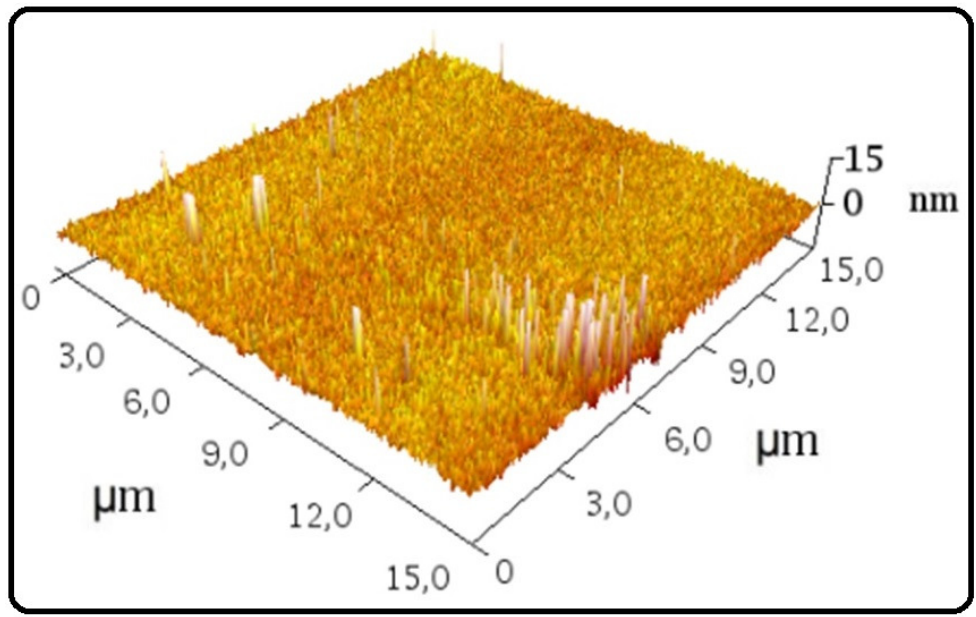

Figure 4: Gonzalez Sutter, et. al.

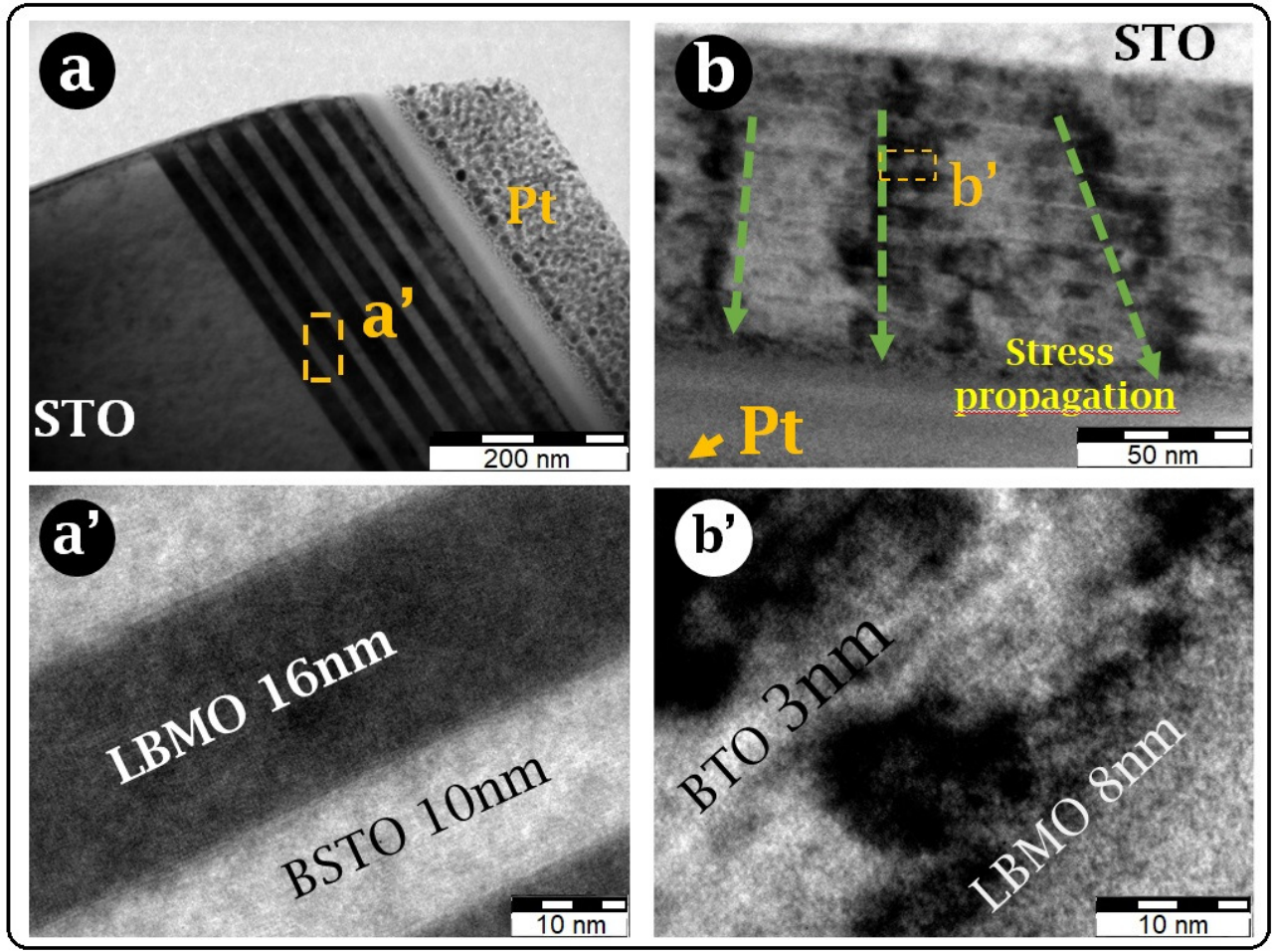

Figure 5: Gonzalez Sutter, et. al.

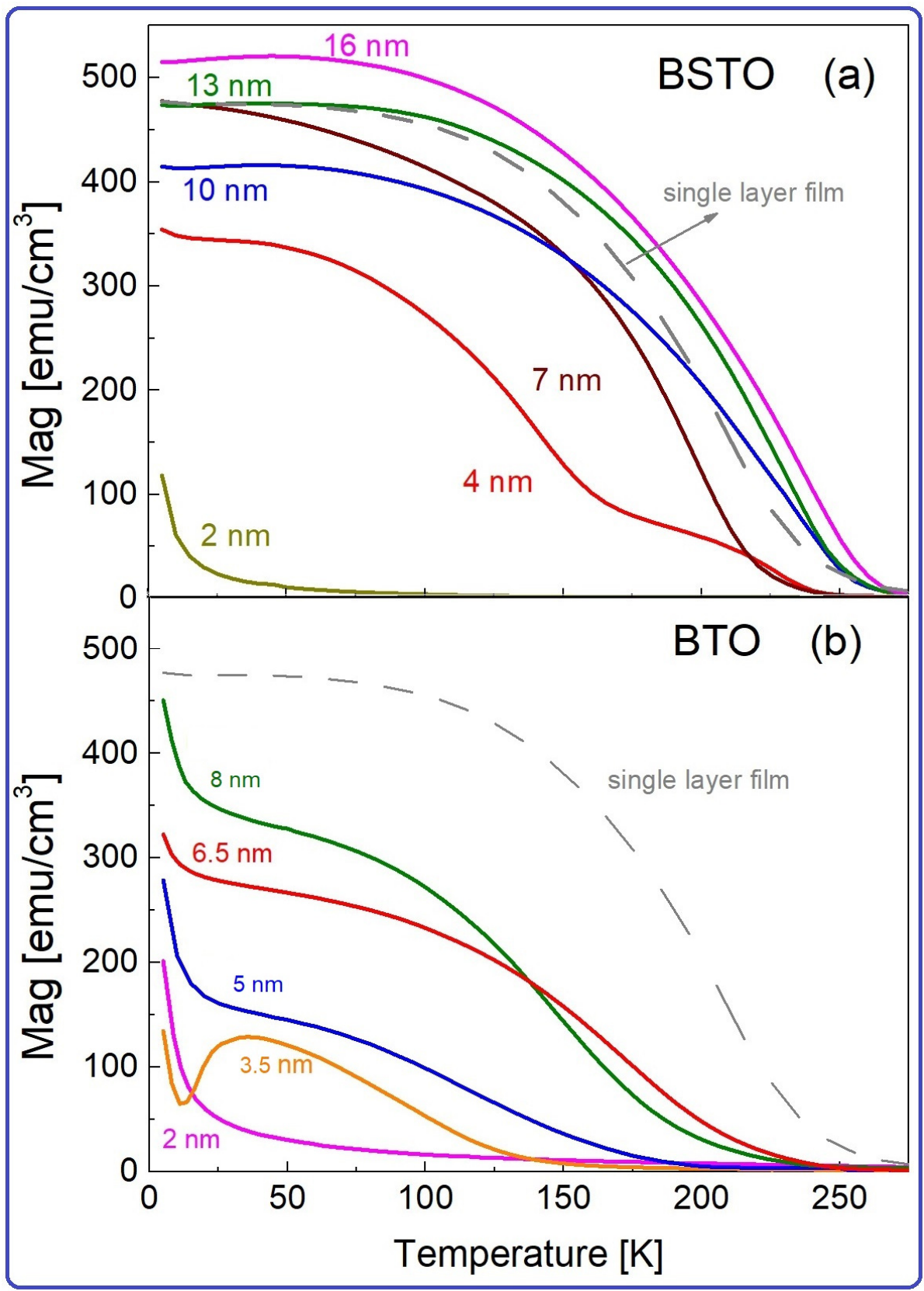

Figure 6: Gozalez Sutter, et. al.

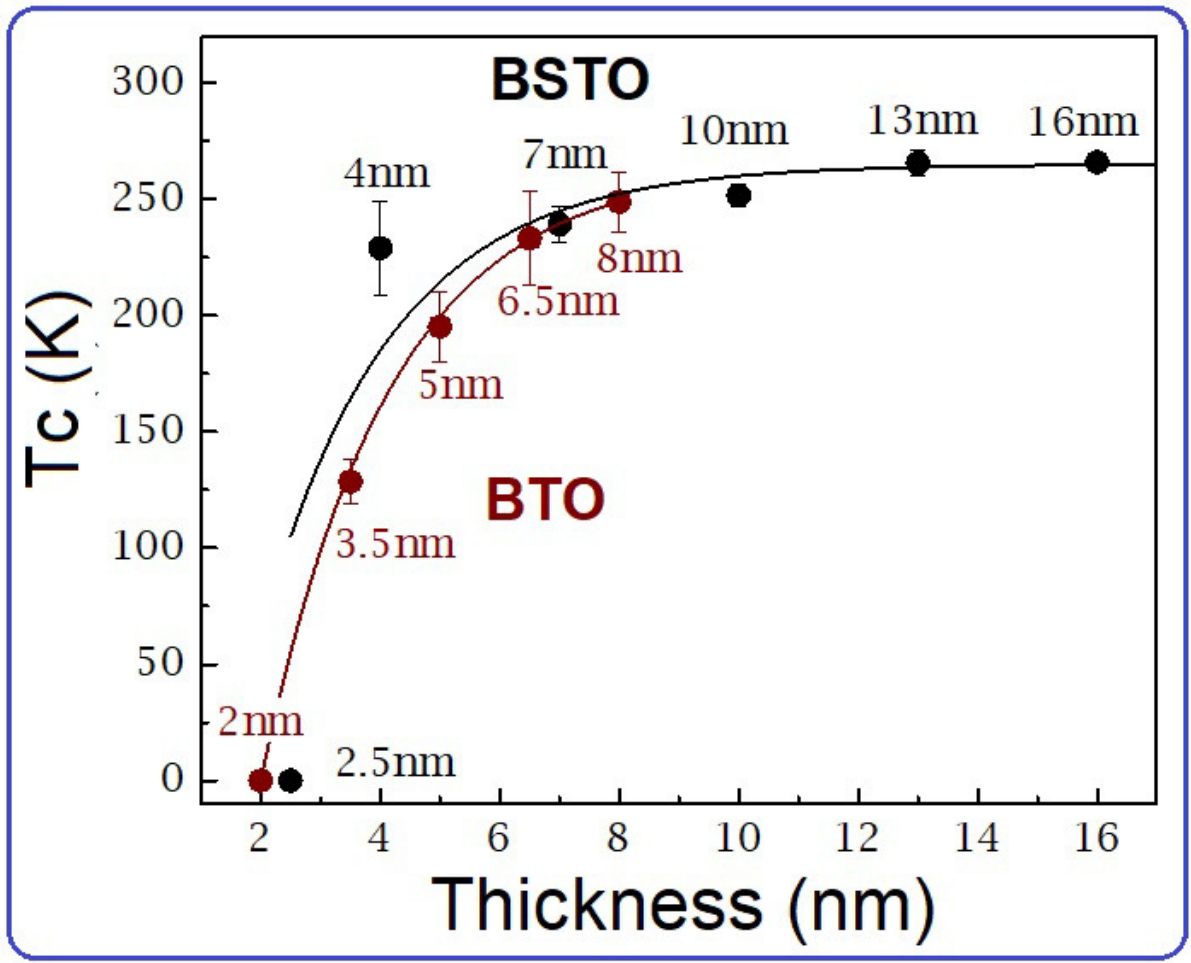

Figure 7: Gonzalez Sutter, et. al.

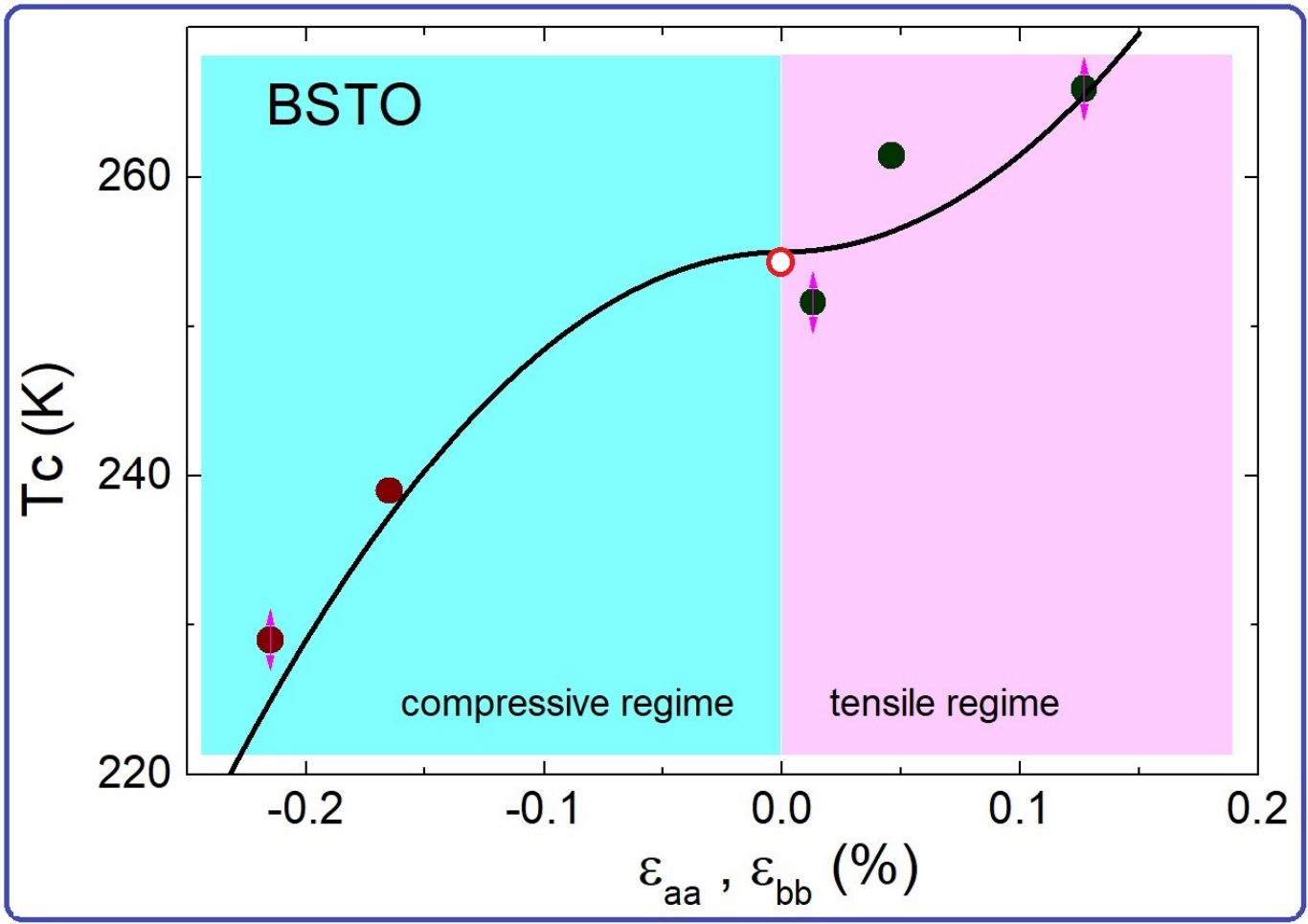

Figure 8. Gonzalez Sutter, et. al.

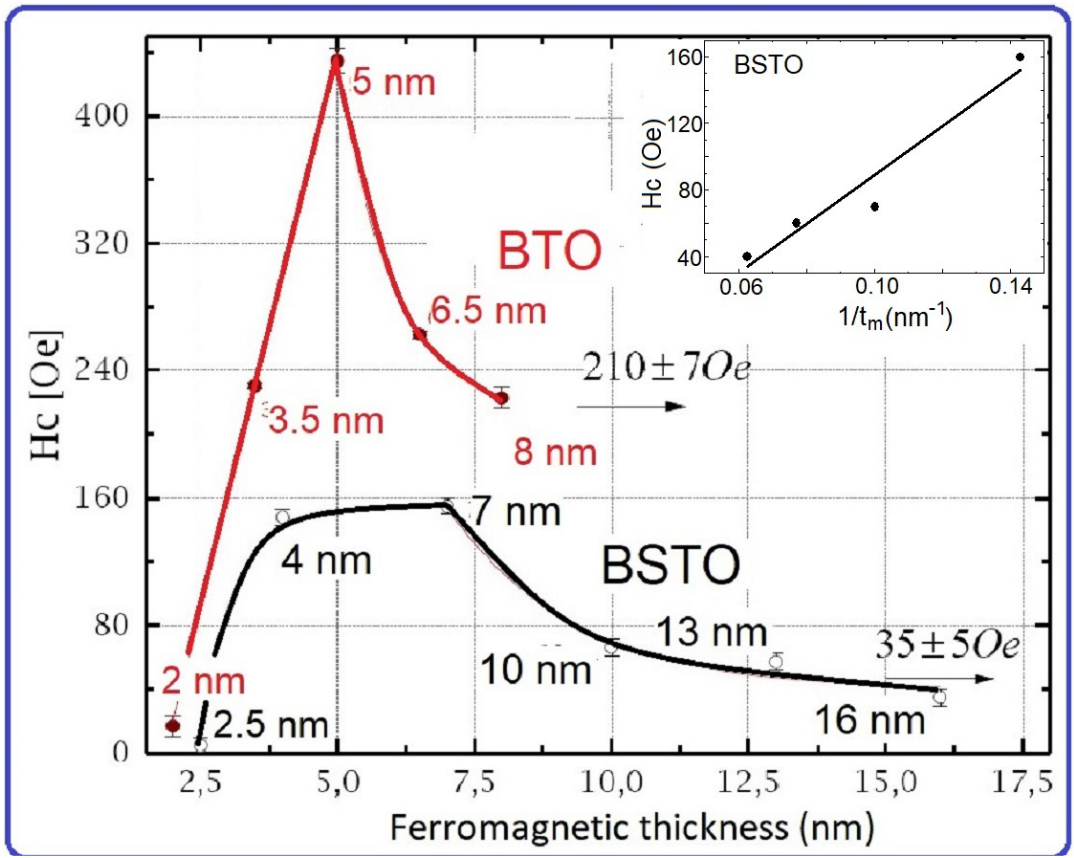

Figure9 Gonzalez Sutter, et. al.

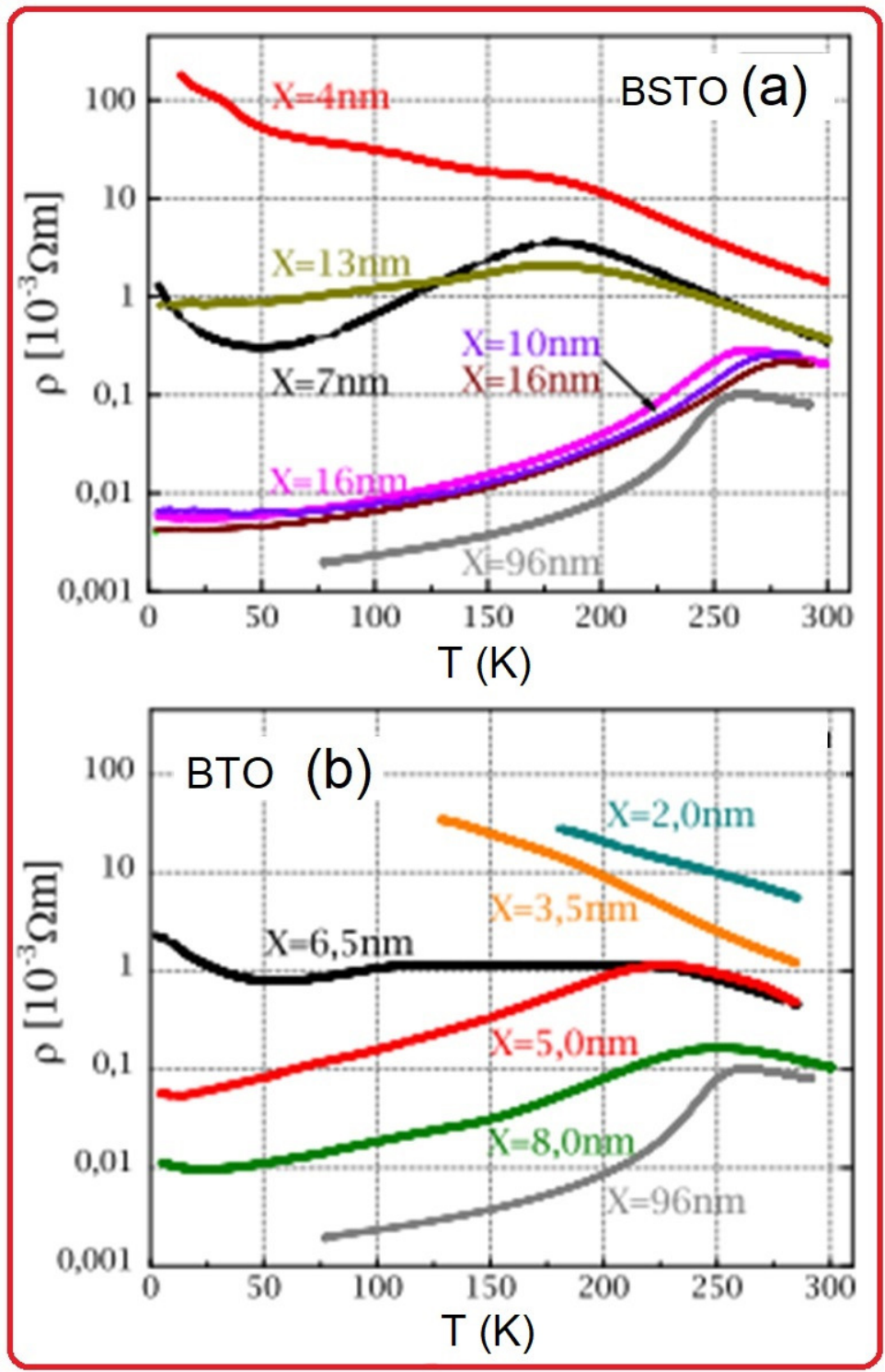

Figure 10: Gonzalez et. al.

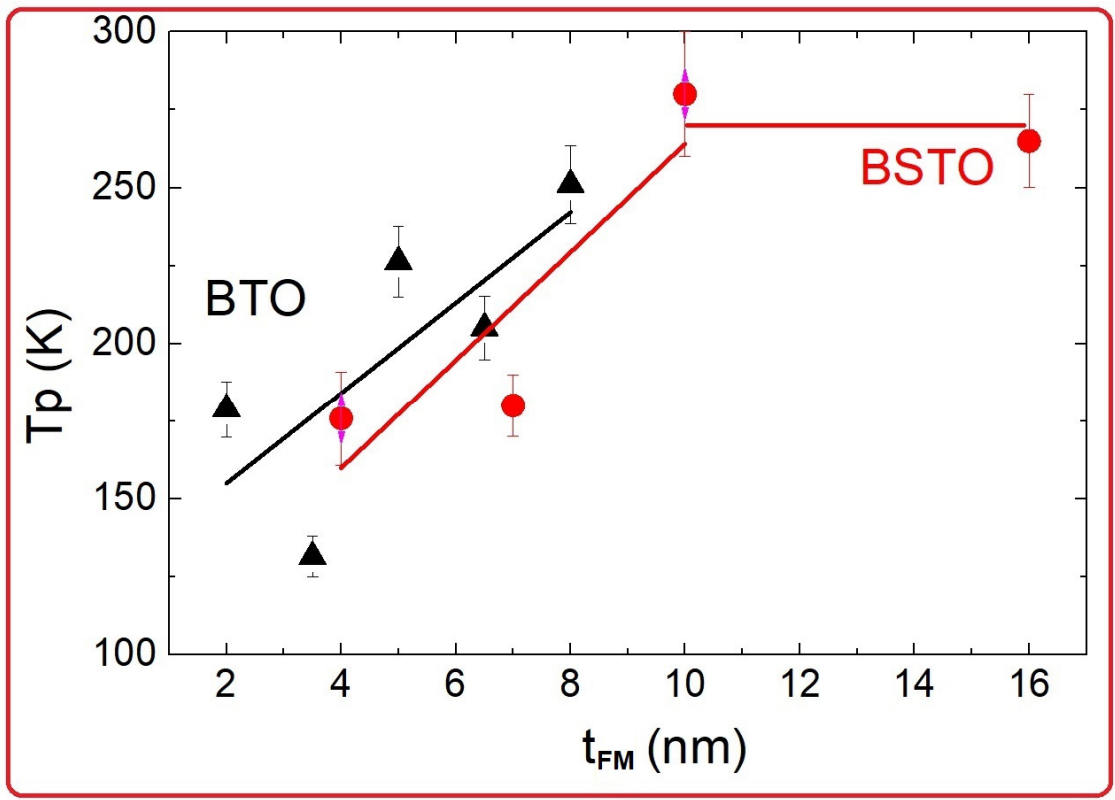

Figure 11: Gozalez Sutter, et. al.

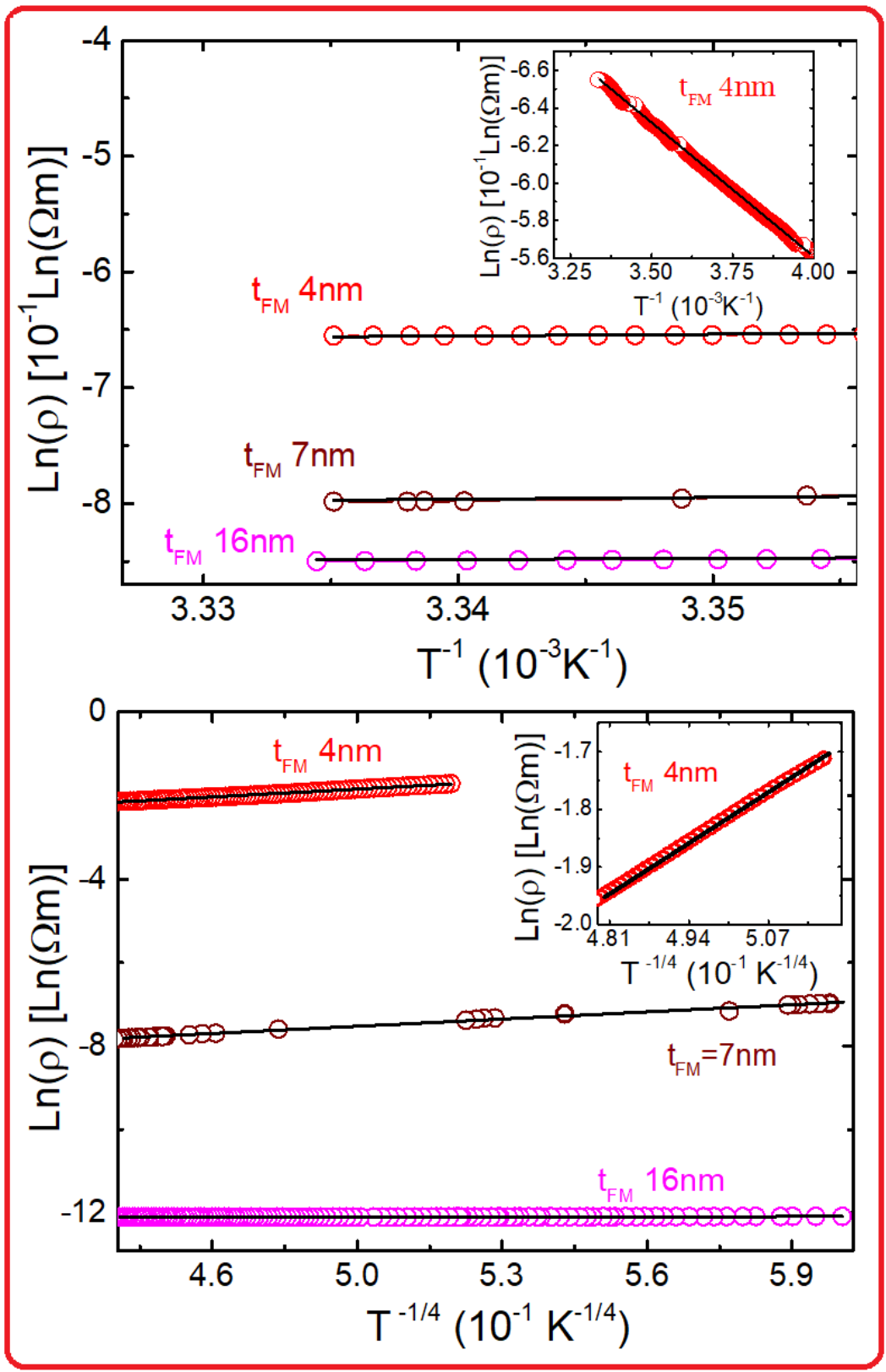

Figure 12 Gozalez Sutter, et. al.

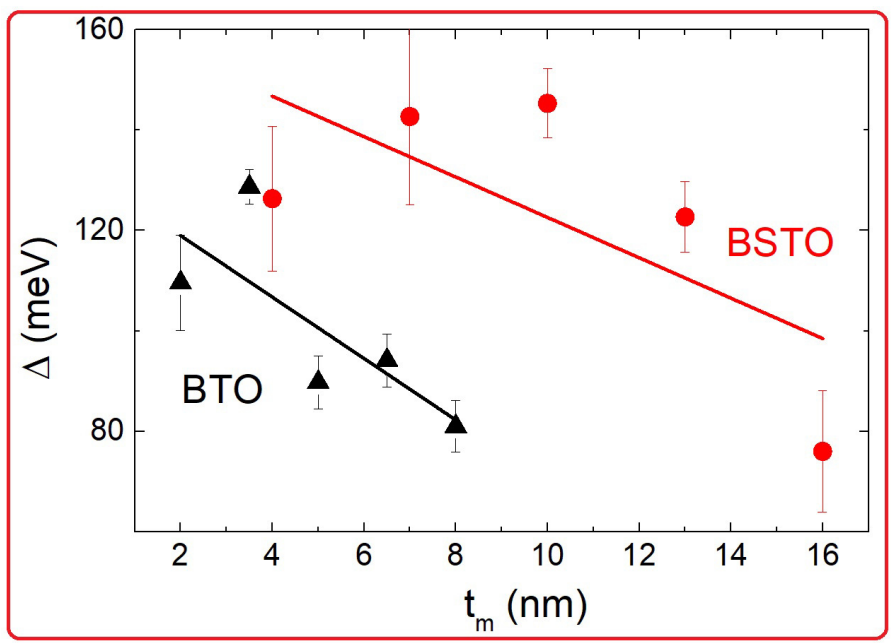

Figure 13: Gonzalez et. al.

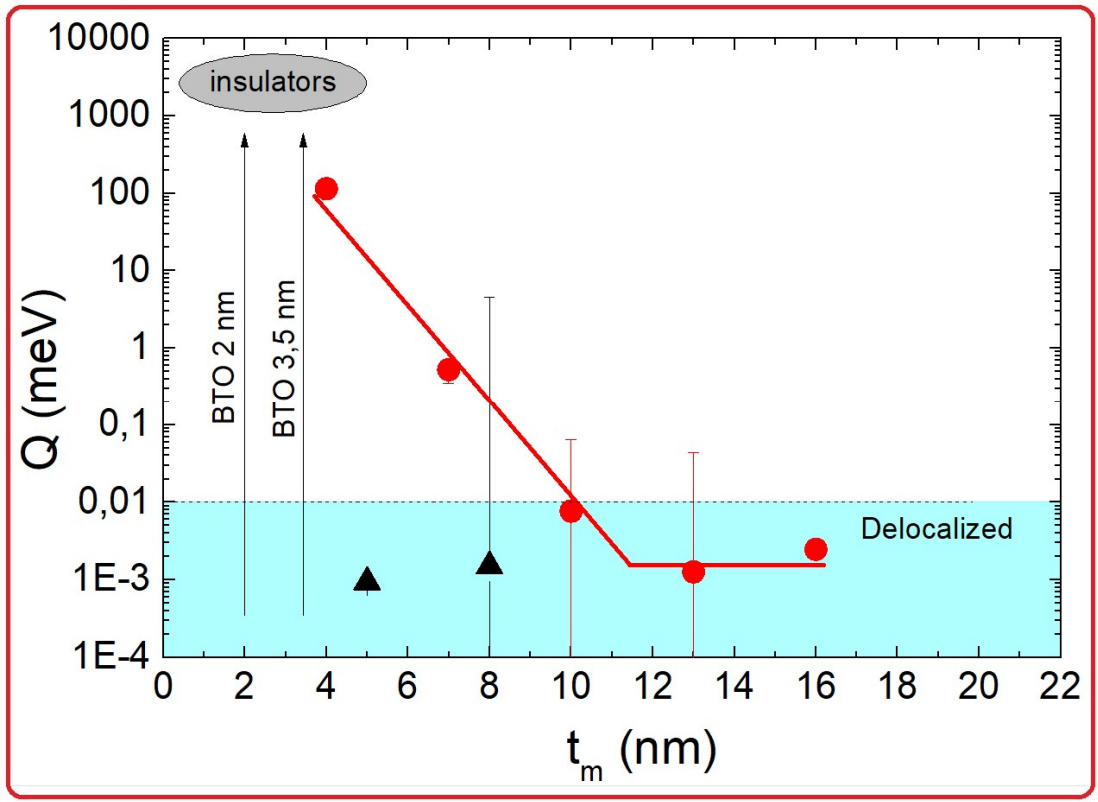